**Transfer Learning from an Artificial Radiograph-landmark Dataset for Registration of the Anatomic Skull Model to Dual Fluoroscopic X-ray Images**

Chaochao Zhou [1, 2], Thomas Cha [1, 2], Yun Peng [1], Guoan Li [1, *]

[1] Orthopaedic Bioengineering Research Center, Newton-Wellesley Hospital and Harvard Medical School, Newton, MA, USA; [2] Department of Orthopaedic Surgery, Massachusetts General Hospital and Harvard Medical School, Boston, MA, USA

* To whom correspondence should be addressed:
Guoan Li, Ph.D.
Orthopaedic Bioengineering Research Center
Newton-Wellesley Hospital
Harvard Medical School
159 Wells Avenue
Newton, MA, 02459, USA
Phone: +1 (617) 530-0563
Email: gli1@partners.org

**Abstract**

Registration of 3D anatomic structures to their 2D dual fluoroscopic X-ray images is a widely used motion tracking technique. However, deep learning implementation is often impeded by a paucity of medical images and ground truths. In this study, we proposed a transfer learning strategy for 3D-to-2D registration using deep neural networks trained from an artificial dataset. Digitally reconstructed radiographs (DRRs) and radiographic skull landmarks were automatically created from craniocervical CT data of a female subject. They were used to train a residual network (ResNet) for landmark detection and a cycle generative adversarial network (GAN) to eliminate the style difference between DRRs and actual X-rays. Landmarks on the X-rays experiencing GAN style translation were detected by the ResNet, and were used in triangulation optimization for 3D-to-2D registration of the skull in actual dual-fluoroscope images (with a non-orthogonal setup, point X-ray sources, image distortions, and partially captured skull regions). The registration accuracy was evaluated in multiple scenarios of craniocervical motions. In walking, learning-based registration for the skull had angular/position errors of $3.9 \pm 2.1°/4.6 \pm 2.2$ mm. However, the accuracy was lower during functional neck activity, due to overly small skull regions imaged on the dual fluoroscopic images at end-range positions. The methodology to strategically augment artificial training data can tackle the complicated skull registration scenario, and has potentials to extend to widespread registration scenarios.

## 1. Introduction

Registration of anatomic models (3D) to dual fluoroscopic X-ray images (2D) is a widely used approach to accurately tracking *in vivo* motions of anatomic bony structures [1] without soft tissue artifacts that were commonly introduced by optical motion capture systems [2]. Clinically, 3D-to-2D registration has key applications in preoperative surgical planning, image-guided surgery, and postoperative evaluation [3–6]. Recently, a manual 3D-to-2D registration approach has been leveraged to investigate craniocervical kinematics [7]. The manual registration is achieved in a virtual dual-fluoroscope system (**Fig. 1**) created by a computer program, in which anatomic 3D models were translated and rotated in six degrees of freedom (DOFs), until their projections matched the osseous outlines/features captured on the dual fluoroscopic images [7]. However, manual registration is extremely laborious and low-efficient. Typically, it requires several hours to accurately register the skull and cervical vertebrae to a single pair of fluoroscopic images. Therefore, it is highly desirable to introduce intelligent algorithms towards automatic 3D-to-2D registration.

In earlier years, optimization-based 3D-to-2D automatic registration approaches incorporating Canny's edge detection [8], outlining [9], or similarity measures [10] were developed to track *in vivo* motions of human knee joints [11–13]. Generally, optimization in these approaches tends to be trapped at local optima because of non-convex objective functions. To obtain the global optimal registration results and mitigate the sensitivity to initializations, it is necessary to provide better initial alignment [12], adopt multiple initializations [14], or formulate more efficient similarity measures [15]. Owing to the advance of deep neural networks and associated large-scale computation frameworks, learning-based approaches have been applied to 3D-to-2D registration [16–18]. Recently, a POINT$^2$ method using tracking and triangulation networks was proposed to address the multi-view 3D-to-2D rigid registration problem [3]. The tracking network based on a Siamese architecture transferred features on digitally reconstructed radiographs (DRRs) to those on X-rays, which were further fed to the triangulation network for point-based registration. Compared to existing learning-based approaches, it was shown that the POINT$^2$ method achieved excellent performance [3]. Therefore, it suggests that deep neural networks can detect feature points on radiographs (on which humans may not even perform well), and that point-based 3D-to-2D registration by triangulation is a more robust registration approach.

Unlike natural images, medical images are commonly less available because of the concern of high radiation exposure (only hundreds of medical images were adopted in reported deep learning implementations as described above), so it largely limits the prediction accuracy of deep neural networks that are greedy for large quantities of data during training. Furthermore, it is less practicable for us to implement existing learning approaches which require a large number of training labels (*i.e.*, the ground truth positions of 3D bones *in vivo*) corresponding to each pair of fluoroscopic images, as manual registration is an extremely time-consuming task as introduced above. However, we anticipated that more intelligent learning strategies are promising solutions to the dilemma. In this study, we proposed a transfer learning framework including a supervised learning for landmark detection and an unsupervised learning for image style translation; both learning modules were trained from an artificial dataset of radiographs and landmarks (that means, they can be automatically forged and expanded). As a test, we attempted the registration of the 3D skull model to dual fluoroscopic X-ray images, in which the skull was not fully captured (*i.e.*, only the mandible and/or occiput were imaged). The feasibility of the framework was evaluated through the registration accuracy in terms of six DOFs of the 3D skull model in multiple functional activities.

## 2. Methods

As an overview (**Fig. 2**), the proposed transfer learning framework for 3D-to-2D registration consists of three main modules, including landmark detection (*Section 2.1*), image style translation (*Section 2.2*), and point-based registration (*Section 2.3*). After the actual dual fluoroscopic X-ray images were preprocessed (*Section 2.4*), they were fed to deep neural networks to perform learning-based 3D-to-2D registration, and the registration accuracy was evaluated according to the performance measures (*Section 2.5*). This study involved use of CT and dynamic fluoroscopic image data of an asymptomatic female subject, which were collected in previously reported experimental studies [7,19].

### *2.1. Artificial Dataset Generation and Landmark Detection*

A 3D anatomic model of the skull was reconstructed from the craniocervical CT volume of a female subject and total $n_{LM} = 33$ landmarks were attached onto the skull model (**Fig. 3**). In particular, there were 13 pairs of symmetric landmarks, as indicated by paired numbers in **Fig. 3**. The craniocervical CT volume data were rendered to grayscale DRRs using a shear-warp ray-casting algorithm which assumes parallel X-ray beams [20] (a coding implementation is available in [21]). Further development was made such that the 3D skull anatomic model and landmarks were projected to the DRR rendering plane companying with ray casting. The resulting craniocervical DRRs as well as their skull masks and image landmarks with different transformations (*e.g.*, rotations, translations, and scaling) were demonstrated in **Fig. 4**.

Based on facial landmark detection for natural images [22], a deep residual network (ResNet) [23,24] with ~11 million trainable parameters was developed to detect landmarks on DRRs. The architecture of the ResNet was presented in **Fig. 5**. To train the ResNet, a dataset of total 9751 DRR-landmark pairs with different skull positions, orientations and sizes were randomly generated and split to a training set (9251 pairs) and a testing set (500 pairs). Within the entire dataset (9751 pairs), 2139 DRRs were automatically skull-segmented via the skull masks (the blue regions in **Fig. 4**). The input image dimension of the ResNet was set to $128 \times 128 \times 1$. Since each landmark was positioned by two image coordinates, the output dimension of the ResNet was 66 (considering the total 33 skull landmarks on DRRs). An Adam optimizer with a learning rate of 0.001 was used for training. To improve optimization convergence, both the input image intensities (range: [0, 255]) and the output landmark coordinates in the field of view (range: [1, 128]) were normalized to [-1, 1].

*2.2. Image Style Translation between X-rays and DRRs*

There were discernable style differences between X-rays captured by actual fluoroscopes and DRRs generated by the ray-casting algorithm. Since we used DRRs to train the ResNet, it was expected to facilitate landmark detection on real X-rays by translating the X-ray style to the DRR style. In this study, unpaired image-to-image translation between X-rays and DRRs was performed using a cycle generative adversarial network (GAN) [25]. To train the cycle GAN, we collected 6716 randomly generated DRRs as described in *Section 2.1*, as well as 6525 X-rays dynamically captured by dual fluoroscopes (30 Hz), when the head of the subject was moving during walking [19] and neck flexion-extension / lateral bending / axial rotation [7].

In our implementation of the cycle GAN, two main modifications have been made. First, the input image dimension in the original cycle GAN was $256 \times 256 \times 3$ [25], but it caused prediction collapse when we translated X-rays to DRRs, because of less available X-rays compared to natural images [25]. This problem was effectively addressed by feeding both X-rays and DRRs with a reduced dimension of $128 \times 128 \times 1$ to the cycle GAN (it also determined the input dimension of the ResNet in *Section 2.1*). Second, we observed that the identity loss function originally adopted in the cycle GAN [25] did not rigorously preserve contents (*i.e.*, the geometry and position of an imaged object), so it was replaced by a content-preserving loss function ($l_{cp}$) [6]:

$$l_{cp} = 1 - \frac{1}{2}\left(\varphi\left(\boldsymbol{I}_{rX}, \boldsymbol{I}_{fD}\right) + \varphi\left(\boldsymbol{I}_{rD}, \boldsymbol{I}_{fX}\right)\right) \qquad \textbf{(Eq. 1)}$$

where $\varphi$ is the zero normalized gradient cross correlation of two images (please refer to its detailed formulation in [6]). $\boldsymbol{I}_{rX}$ and $\boldsymbol{I}_{rD}$ were two unpaired real images of X-ray and DRR in each batch training (a batch size of 1, *i.e.*, instance normalization was adopted [26]), during which two fake images of DRR and X-ray, $\boldsymbol{I}_{fD}$ and $\boldsymbol{I}_{fX}$, were predicted by the forward and backward generators of the cycle GAN, respectively. $l_{cp}$ has a range between 0 and 1; a smaller value represents higher similarity in the image contents before and after style translation.

*2.3. Point-based 3D-to-2D Registration by Triangulation Optimization*

To search rigid transformations of the 3D model including three rotations ($\boldsymbol{\theta}^*$) and three translations ($\boldsymbol{\tau}^*$) in 3D space, the point-based registration of the 3D skull model to the X-rays of dual fluoroscopes (denoted by F1 and F2, respectively) can be simply described by an optimization problem with an unconstrained objective function ($\mu$) in terms of the Euclidian distances between the sets of predicted and projected landmarks [27]:

$$[\boldsymbol{\theta}^*, \boldsymbol{\tau}^*] = \arg\min_{[\boldsymbol{\theta},\boldsymbol{\tau}]}: \quad \mu(\boldsymbol{\theta}, \boldsymbol{\tau}) = \|\boldsymbol{U}^{F1} - \boldsymbol{V}^{F1}(\boldsymbol{\theta}, \boldsymbol{\tau})\|_F + \|\boldsymbol{U}^{F2} - \boldsymbol{V}^{F2}(\boldsymbol{\theta}, \boldsymbol{\tau})\|_F \qquad \textbf{(Eq. 2)}$$

where $\|\boldsymbol{A}\|_F = \sqrt{\mathrm{trace}(\boldsymbol{A}^T\boldsymbol{A})}$ is the Frobenius norm of a matrix. The global coordinate system was set at the center of the F1 intensifier (**Fig. 1**). $U_{ij}^{F1}$ and $U_{ij}^{F2}$ are the predicted landmark coordinates (note that they have been converted from image coordinates to spatial coordinates following the X-ray image preprocessing steps) on the F1 and F2 intensifiers, respectively. $V_{ij}^{F1}$ and $V_{ij}^{F2}$ are the coordinates projected from landmarks attached on the 3D skull model to the F1 and F2 intensifiers, respectively. For all the coordinate matrices, $j = x, y, z$ denotes each spatial coordinate component; $i = 1, 2, \cdots, n_{vis}$, where $n_{vis}$ is the number of predicted landmarks ($U_{ij}^{F1}$ and $U_{ij}^{F2}$) that are simultaneously visible within both the fields of view of F1 and F2, thus $n_{vis} \leq n_{LM} = 33$. The six DOFs of the 3D skull model relative to the global coordinate system consist of three Euler angles (defined by extrinsic rotations with a sequence of "$zyx$" [28]), $\boldsymbol{\theta} = [\theta_x, \theta_y, \theta_z]$ and three spatial translations with respect to the global coordinate origin, $\boldsymbol{\tau} = [\tau_x, \tau_y, \tau_z]$. Therefore, the optimization is to seek six DOFs ($\boldsymbol{\theta}^*$ and $\boldsymbol{\tau}^*$) of the skull model, such that the differences of $V_{ij}^{F1}(\boldsymbol{\theta}, \boldsymbol{\tau})$ from $U_{ij}^{F1}$ and of $V_{ij}^{F2}(\boldsymbol{\theta}, \boldsymbol{\tau})$ from $U_{ij}^{F2}$ are minimized simultaneously. For all optimizations, the optimization variables were always initialized at $\boldsymbol{\theta} = [0,0,0]$ and $\boldsymbol{\tau} = \boldsymbol{\tau}^0$, where $\boldsymbol{\tau}^0$ is the center of the virtual dual-fluoroscope system, *i.e.*, the average coordinates of F1 and F2 sources and intensifiers.

It is noted that the attenuations of X-ray images mainly depend on bone density, causing a difficulty in distinguishing objects close or distant to the X-ray source. For example, as shown in **Fig. 6**, the left and right mandibles of the subject cannot be distinguished on X-rays. The only way to distinguish them is by anatomic features; coincidently, the subject has an abnormal wisdom tooth on the left lower jaw (**Fig. 6**). This is in contrast to DRRs, in which close and distant objects appear to have different attenuations. Since we used DRRs to train the ResNet for landmark detection, the predicted skull landmarks for real X-rays (and fake DRRs) could be mirrored (recall that there were 13 pairs of symmetric skull landmarks as shown in **Fig. 2**). Therefore, in point-based registration, the optimization (**Eq. 2**) needed to be run four times, with a strategy to exchange the coordinates of the predicted symmetric landmarks ($U_{ij}^{F1}$ and $U_{ij}^{F2}$) on the F1 and F2 intensifiers (**Table 1**). The six DOFs of the skull were ultimately chosen to be those ($\hat{\boldsymbol{\theta}}$ and $\hat{\boldsymbol{\tau}}$) corresponding to the minimum of the optimal objective function values after the four optimizations.

### *2.4. Preprocessing of Fluoroscopic X-ray Images in 3D-to-2D Registration*

After training both the ResNet and the cycle GAN, in total 48 pairs of dynamic craniocervical dual fluoroscopic images of the subject during walking [19] and neck flexion-extension / lateral bending / axial rotation [7] (*i.e.*, 12 pairs in each scenario) were chosen to perform both manual and learning-based registration of the 3D skull model in the virtual dual-fluoroscope system (**Fig. 1**). As style translation may generate image distortion occurred outside the skull region, it would mislead the recognition of the skull region by the ResNet. Therefore, prior to the learning-based registration, we manually segmented the skulls on real X-rays, as illustrated in **Fig. 7a**. Since the skull region was highly preserved by the content-preserving loss function (**Eq. 1**) during style translation, we further segmented the skulls on the corresponding DRRs after style translation using the skull-segmented real X-ray as masks (**Fig. 7a**).

Different from DRRs, actual fluoroscopic images were generated by point X-ray sources and typically distorted because of the use of image intensifiers [29]. Hence, the preprocessing of X-ray images was required to establish a virtual dual-fluoroscope system for 3D-to-2D registration (**Fig. 1**) [12]. First, image distortions on each individual X-ray were corrected by an acrylic calibration plate consisting of stainless steel bead arrays in a regular space (**Fig. 7b**) and the deformation field was fitted using a fifth-order polynomial [29]. Furthermore, using a source alignment tool with four implanted stainless steel beads, the relative position of dual fluoroscopes in an experimental setup (noting that the dual fluoroscopes were not aligned perfectly orthogonally to each other) was determined by optimization (**Fig. 7c**). Correspondingly, the predicted skull landmark coordinates (detected by the ResNet) on fake DRRs (after style translation from real X-rays) also needed to experience the image distortion correction transform, and be aligned to the intensifier planes considering the actual layout of the dual fluoroscopes.

### *2.5. Evaluation of Point-based Registration Accuracy*

In terms of tracking bone motion *in vivo*, the exact bone positions are unknown, so we benchmarked the point-based registration in the proposed deep learning framework against manual registration performed by human operators [7]. Using cadaveric specimens with implanted beads, manual registration has been validated to be a reliable approach to reproduce cervical kinematics [30]. Therefore, in this study, manually registered model DOFs were used to represent the ground truths.

The manual registration of each pair of fluoroscopic images took an estimated duration of 1~2 hours depending on the head near neutral (easier) or end-range (harder) positions, such that the projections of the skull model onto F1 and F2 intensifiers were tuned to have maximal intersection-over-union with respect to the skull regions on both X-rays. The angular ($\varepsilon_\theta$) and position ($\varepsilon_\tau$) errors of the point-based registration with regard to the manual registration were defined, respectively:

$$\varepsilon_\theta = \left\|\boldsymbol{\theta}^M - \widehat{\boldsymbol{\theta}}\right\|_\infty$$
$$\varepsilon_\tau = \left\|\boldsymbol{\tau}^M - \widehat{\boldsymbol{\tau}}\right\|_\infty \quad \textbf{(Eq. 3)}$$

where $\|\boldsymbol{v}\|_\infty = \max_i |v_i|$ is the infinity norm of a vector. $\boldsymbol{\theta}^M$ and $\boldsymbol{\tau}^M$ are the six DOFs of the 3D skull model achieved by manual registration. $\widehat{\boldsymbol{\theta}}$ and $\widehat{\boldsymbol{\tau}}$ are the six DOFs of the 3D skull model achieved by the point-based registration using optimization (*Section 2.3*).

## 3. Results

### *3.1. Performance of ResNet Predictions*

The total number of epochs was set as 300 for training the ResNet. The performance metrics to evaluate the predictions for both the training and testing sets were chosen to be the mean square error (MSE). The training experienced ~10 minutes on a GPU with a RAM of 25 GB and stopped at epoch 134 because there was no further improvement of the ResNet loss. After training, the logarithm base 10 of the MSEs of the predictions for the training and testing sets were -4.86 and -2.96 (in terms of the normalized landmark coordinates), respectively. The landmark predictions and labels in the testing set were visualized in **Fig. 8**, showing an outstanding capability of landmark detection for DRRs with/without skull segmentation.

### *3.2. Performance of Cycle GAN Predictions*

The cycle GAN was trained for 40 epochs (~7.6 hours on a GPU with a RAM of 25 GB). After 30 epochs, no distinct changes were observed on the style-translated images. Both the forward (real X-rays to fake DRRs) and backward (real DRRs to fake X-rays) style translations made by two respective generators in the cycle GAN were demonstrated in **Fig. 9**. It can be observed that the skull region was well persevered in both translations owing to the content-preserving loss function. In terms of the DRR style, the difference between fake and real DRRs was almost indiscernible visually (**Fig. 9**).

### *3.3. Performance of Point-based Registration*

Predicted landmarks on radiographs in different motion scenarios were used in point-based registration; in each registration, four optimizations were performed within 1 second. The accuracies in point-based registration using landmarks predicted from the skull-segmented real X-rays and the corresponding fake DRRs were compared (**Fig. 7a**). Taking manually registered six DOFs of the 3D skull model as the benchmark, the quantitative evaluation of point-based registration in different scenarios were shown in **Fig. 10**. Overall, both angular and position accuracies of registration using landmarks predicted from fake DRRs was at least two-fold superior to those using landmarks predicted from real X-rays. In particular, learning-based registration using fake DRRs in walking showed a promising accuracy, with angular/position errors of $3.9 \pm 2.1°/4.6 \pm 2.2$ mm (**Fig. 10**). The registration results to track head motion during walking were graphically presented in **Fig. 11**; for the learning-based registration using real X-rays, there was obvious misalignment of the 3D skull model projections with the radiographic skull outlines on both F1 and F2 intensifiers. However, the learning-based registration accuracy using fake DRRs were poorer during neck flexion-extension ($8.9 \pm 3.6°/11.9 \pm 6.5$ mm), lateral bending ($14.1 \pm 6.2°/12.4 \pm 7.4$ mm), and axial rotation ($8.9 \pm 4.0°/8.0 \pm 3.9$ mm, **Fig. 10**), as a result of small skull regions on dual fluoroscopic images at end-range positions (**Figs. A1-A3** in *Appendix A*).

## 4. Discussion

In the transfer learning framework, we introduced a DRR-landmark dataset for data augmentation in the training of the ResNet for landmark detection, and for style transfer using the cycle GAN to eliminate the difference between DRR and X-ray. Using the framework, we tackled a challenging registration problem that partial skull regions were imaged in craniocervical dual fluoroscopic X-rays, and evaluated registration accuracy in a variety of head movements, instead of only considering ideal poses. Our testing results showed that the registration accuracy was higher in walking than those in neck flexion-extension, lateral bending and axial rotation, because only a small portion of the skull was visualized in the fields of view of intensifiers at end-range positions during these neck motions (**Figs. A1-A3** in *Appendix A*). Furthermore, the registration accuracy should be conservative, as we did not introduce any fake DRRs (after style translation from real X-rays) and manually registered landmark labels to train the ResNet. Therefore, it demonstrates that our strategy of transfer learning from artificial datasets is feasible, and can help implement deep learning when medical images are scarce and ground truths are difficult to establish. It is also promising to extend this framework to kinematic investigations of other human joints. Each module in the framework played an important role and is discussed below.

*4.1. Landmark Detection*

We decomposed the multi-view registration problem to single-view landmark detection tasks. This largely facilitated deep learning, as the training examples were doubled. Moreover, for single-view landmark detection, we do not need to consider the actual layout of dual fluoroscopes, so the trained ResNet can be applied to multi-view registrations with different experimental settings. In this study, we implemented the shear-warp ray-casting algorithm to generate DRRs; compared to other algorithms, shear-warp ray casting is computationally efficient [20], so it enables us to rapidly expand the training dataset (~1 second per DRR). Although parallel-beam (DRRs) and fan-beam (fluoroscopic X-rays) ray casting typically leads to different rendering geometries, we demonstrated that landmark detection is less sensitive to the type of ray casting. This is not surprising, as landmark detection relies on outlines and features of bony structures on radiographs that deep convolutional networks excel in perceiving [31].

*4.2. Image Style Translation*

Unfortunately, landmark detection is very sensitive to image styles, so the ResNet trained from DRRs cannot be directly applied to X-rays. Theoretically, the ray-casting mapping from CT Hounsfield Unit values to DRR intensities can be calibrated to match the intensity at each pixel on the X-ray, but paired DRRs and X-rays do not exist. Moreover, X-ray intensities vary across different fluoroscope modalities, so a high-fidelity ray-casting algorithm is always less generalizable to other modalities. Previously, we have attempted image intensity histogram equalization [3] as simple style translation between X-rays and DRRs, but the registration accuracy was little improved. Therefore, we introduced the cycle GAN [25] for translation of unpaired X-rays and DRRs. It is shown to be an essential module in our framework, as the registration accuracy using landmarks predicted from fake DRRs are markedly superior to that from real X-rays (**Fig. 10**). Compared to the previous implementation in knee radiographs at full-extension positions [6], we achieved more complex style translation for craniocervical radiographs in various motion scenarios with high preserved contents (**Fig. 9**), by the relatively large training dataset of X-rays and DRRs.

*4.3. Point-based Registration*

We reinforced the notion that point-based registration is robust and insensitive to initial conditions, due to the convex objective function in optimization, in contrast to edge- / outlining- / similarity measure-based registration which potentially requires additional manual manipulations. For the fluoroscope modality that we adopted, the attenuations on X-rays caused a difficulty in the determination of the orientation of a symmetric object. It is a primary challenge to human operators during manual registrations. A successful registration requires a human operator to repeatedly correct the alignment of 3D models until the model projections are matched to radiographic outlines/features on both dual fluoroscopic images. For deep neural networks, correspondingly, symmetric landmarks predicted on fake DRRs (after style translation from X-rays) may be mistakenly mirrored. This problem was well overcome by running four optimizations in each point-based registration. In addition, it should be noted that the registration accuracy in terms of six DOFs of a 3D model is determined by all available predicted landmarks (individual landmarks are not decisive unless there are remarkable biases). Therefore, it is important to predict sufficient landmarks, such that more landmarks can occur in the fields of view of both intensifiers.

*4.4. Limitations and Future Work*

Several limitations need to be addressed to improve registration accuracy and generalize the deep learning implementation. DRRs were generated based on the CT volume in the supine position without intervertebral relative motions as occurring in functional activity, so it appears that the neck in the DRRs is always straight (**Fig. 4**). It potentially increases the difficulty of the cycle GAN in style translation for X-rays captured in actual activity. Correspondingly, image distortion outside the skull regions may occur and affect the registration accuracy, and manual segmentation of the skull region in the actual X-ray images was required to preclude these distorted regions. It is anticipated that only a single component in an image can facilitate style translation and landmark detection, so automatic segmentation should be implemented by combining YOLO object detection [32] and image-to-image (paired) style translation [26]. Moreover, other domain adaption methods [33] should be also attempted to compare with the cycle GAN in terms of robustness. It should be acknowledged that only a single subject was tested in this feasibility study. However, for more subjects, point-to-point correspondences of 3D landmarks between subjects are required. Consistent 3D landmarks can be mapped between subjects according to the deformation field of a statistical shape model [34]. Furthermore, radiation reduction is attractive in clinical practice, as the widely used CT modalities for anatomic reconstruction require high radiation exposure. The transfer learning framework can be further developed for 3D reconstruction by incorporating statistical shape modeling [35]. Furthermore, end-to-end domain adaptation implementations for 3D reconstruction have emerged [6,36], but the learning from the dual fluoroscopic images with actual setups (*i.e.*, the non-orthogonal layout), point X-ray sources, and image distortions still needs further development.

## 5. Conclusion

A transfer learning strategy including landmark detection, style translation, and point-based registration was proposed for 3D-to-2D registration. A DRR-landmark dataset was automatically created for data augmentation in the training of a ResNet for landmark detection, and the style difference between DRR and X-ray was eliminated by style translation using the cycle GAN. It is shown that the proposed strategy is feasible to tackle the registration of the skull model to dual fluoroscopic images where the skull was not completely captured. The strategy for 3D-to-2D registration can be extended to tracking motions of a wide variety of human joints, and further refinement is essential to achieve better performance.

**Acknowledgements**

We are thankful for the support from National Institutes of Health (1R03AG056897), USA.

**Conflict of Interest**

The authors declare that there is no conflict of interest.

**Supplementary Material**

Appendix A: Graphic Presentation of Registration of the 3D Skull Model in Neck Functional Motions

**Figures**

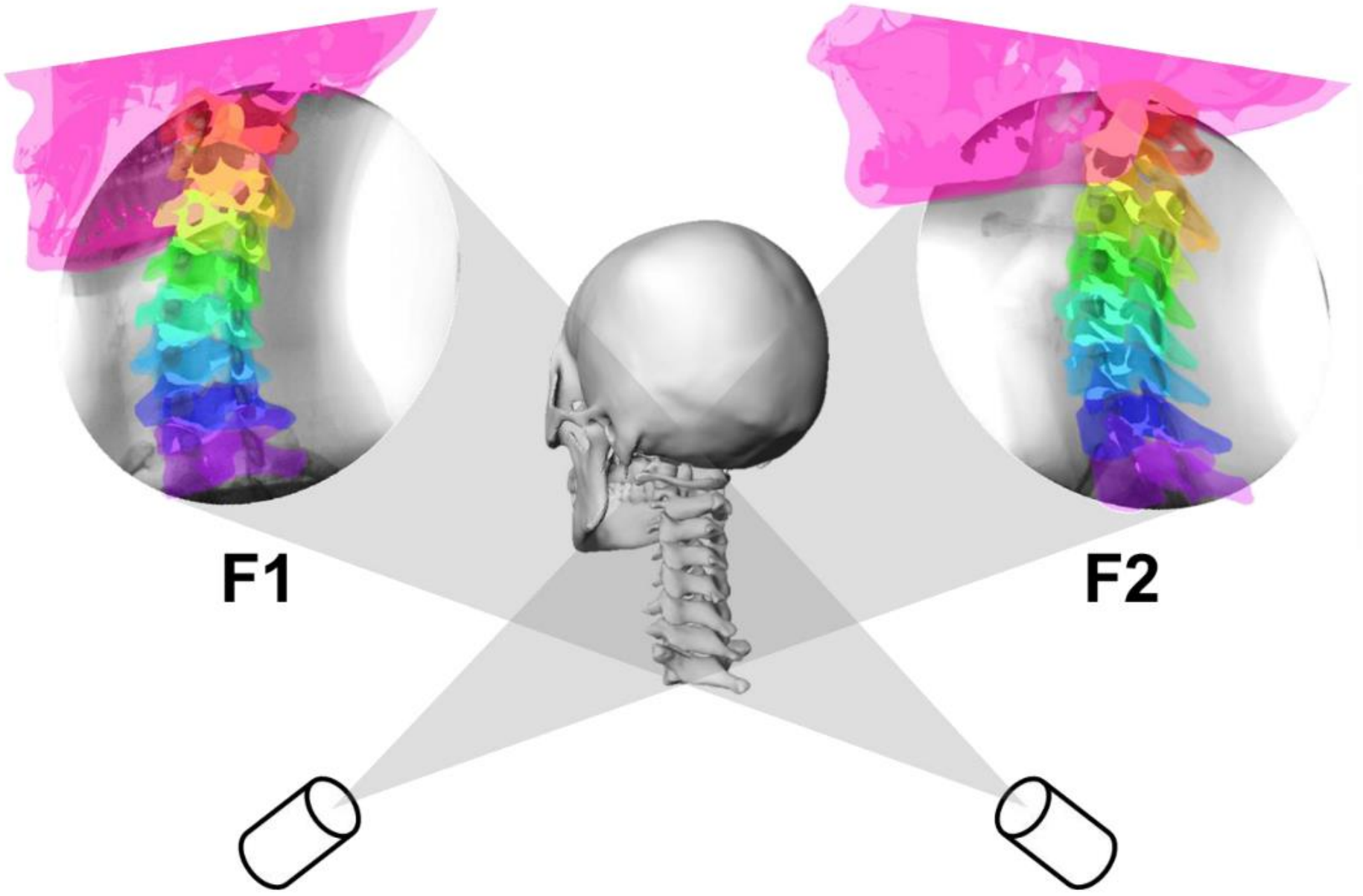

**Fig. 1**: Illustration of manual 3D-to-2D registration operated in a virtual dual-fluoroscope system (*F* = fluoroscope). The color areas on both fluoroscopic images represent the projections of the 3D skull and cervical vertebral models.

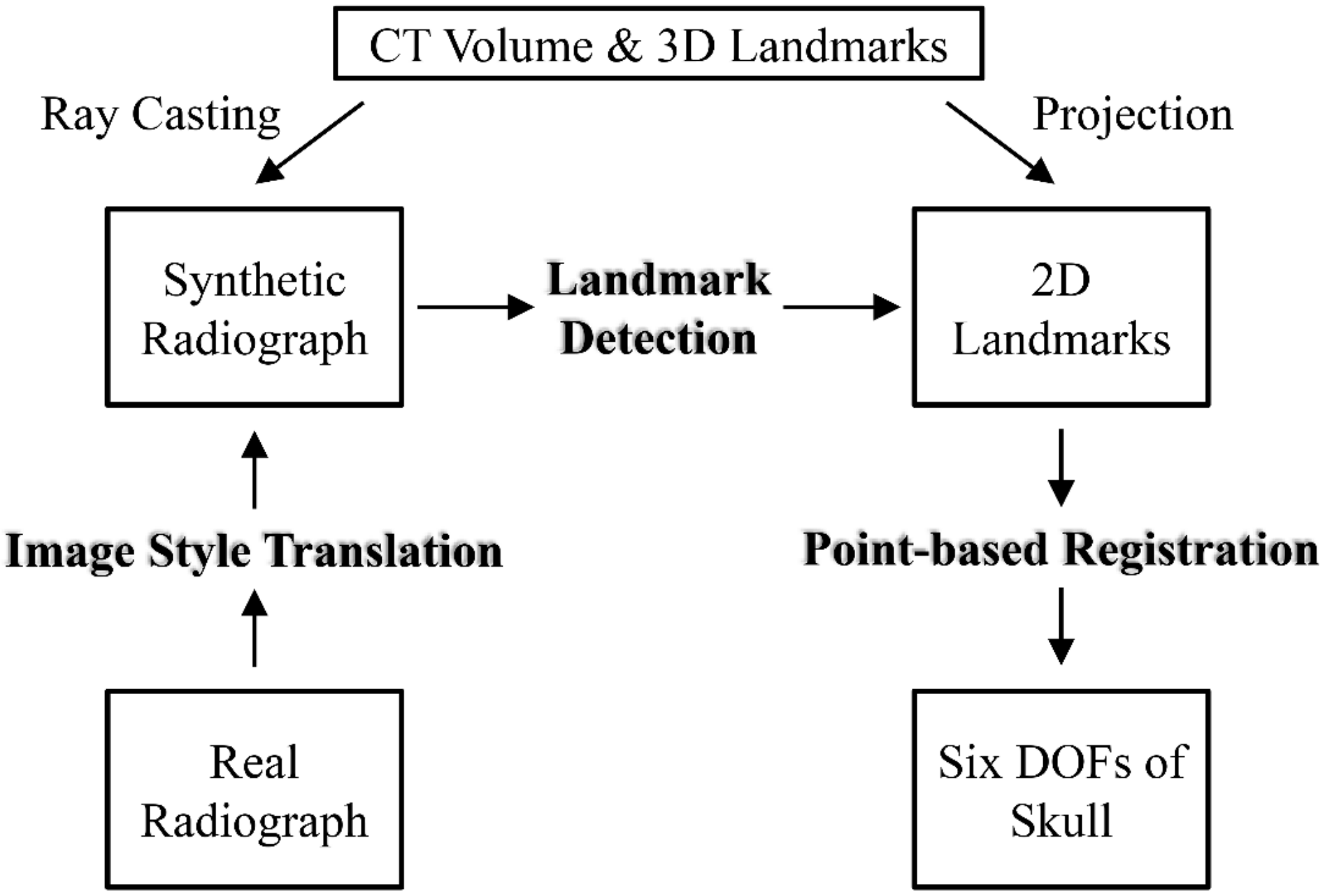

**Fig. 2**: A flow chart of this transfer learning framework, including landmark detection, image style translation, and point-based registration.

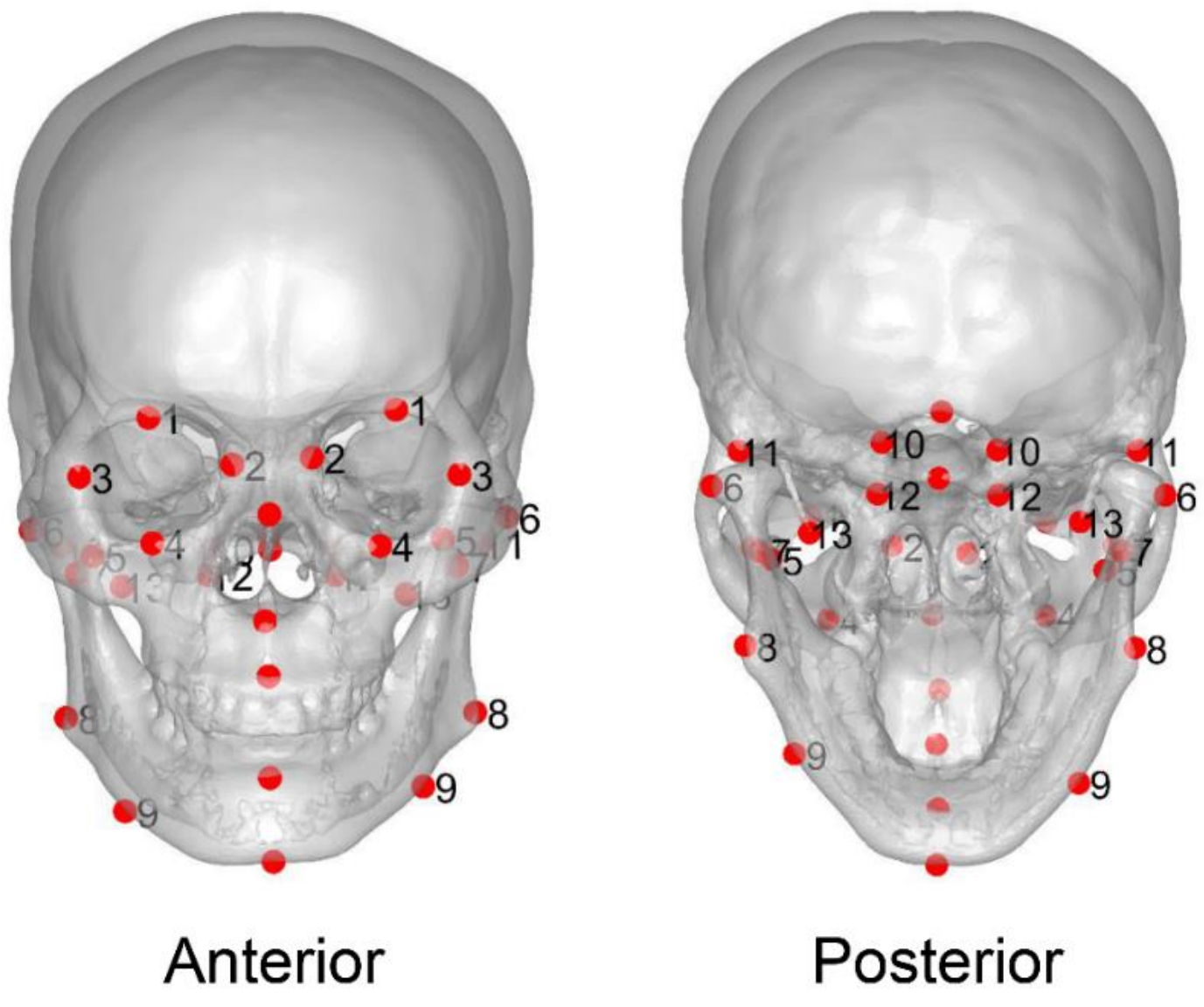

**Fig. 3**: The anatomic landmarks on the 3D skull models. The symmetric landmarks were indicated by paired numbers.

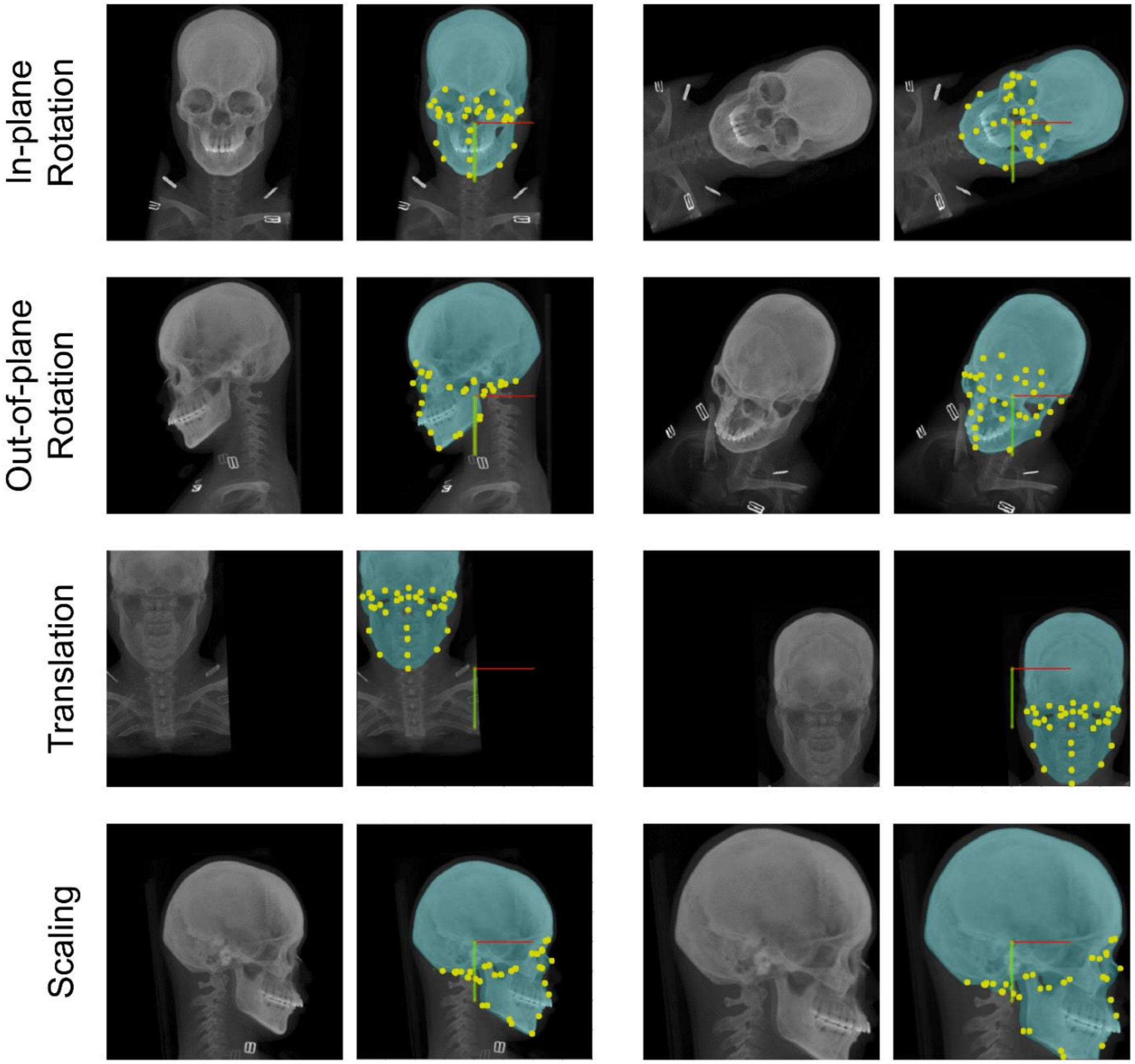

**Fig. 4**: Transformations (rotations, translations, and scaling) of DRRs and their corresponding skull masks (*blue regions*) and image landmarks (*yellow points*).

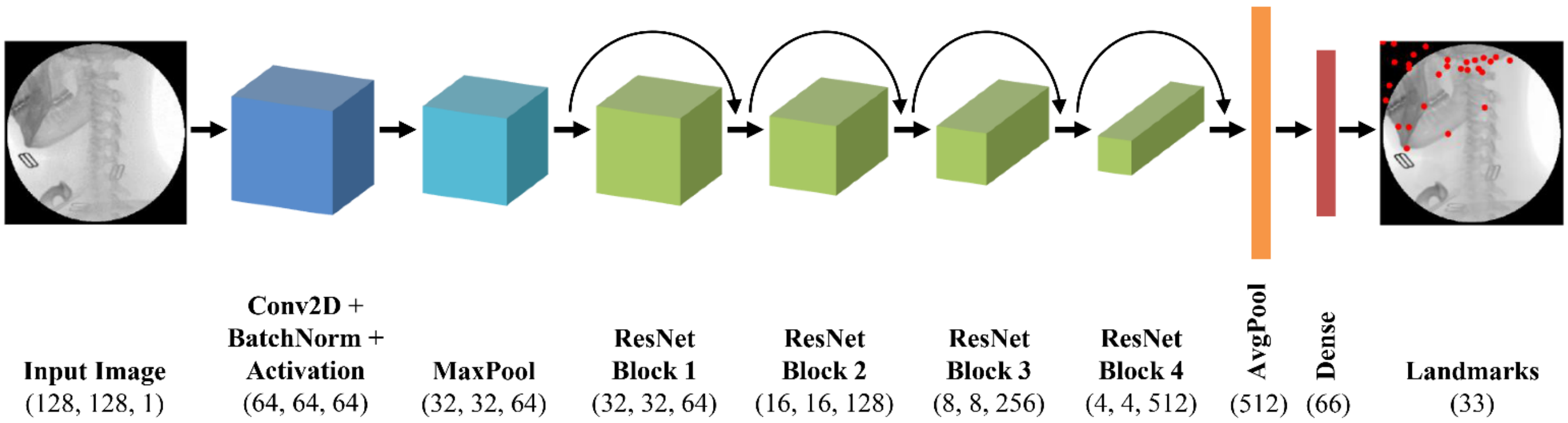

**Fig. 5**: The architecture of the ResNet used to detect landmarks on DRRs.

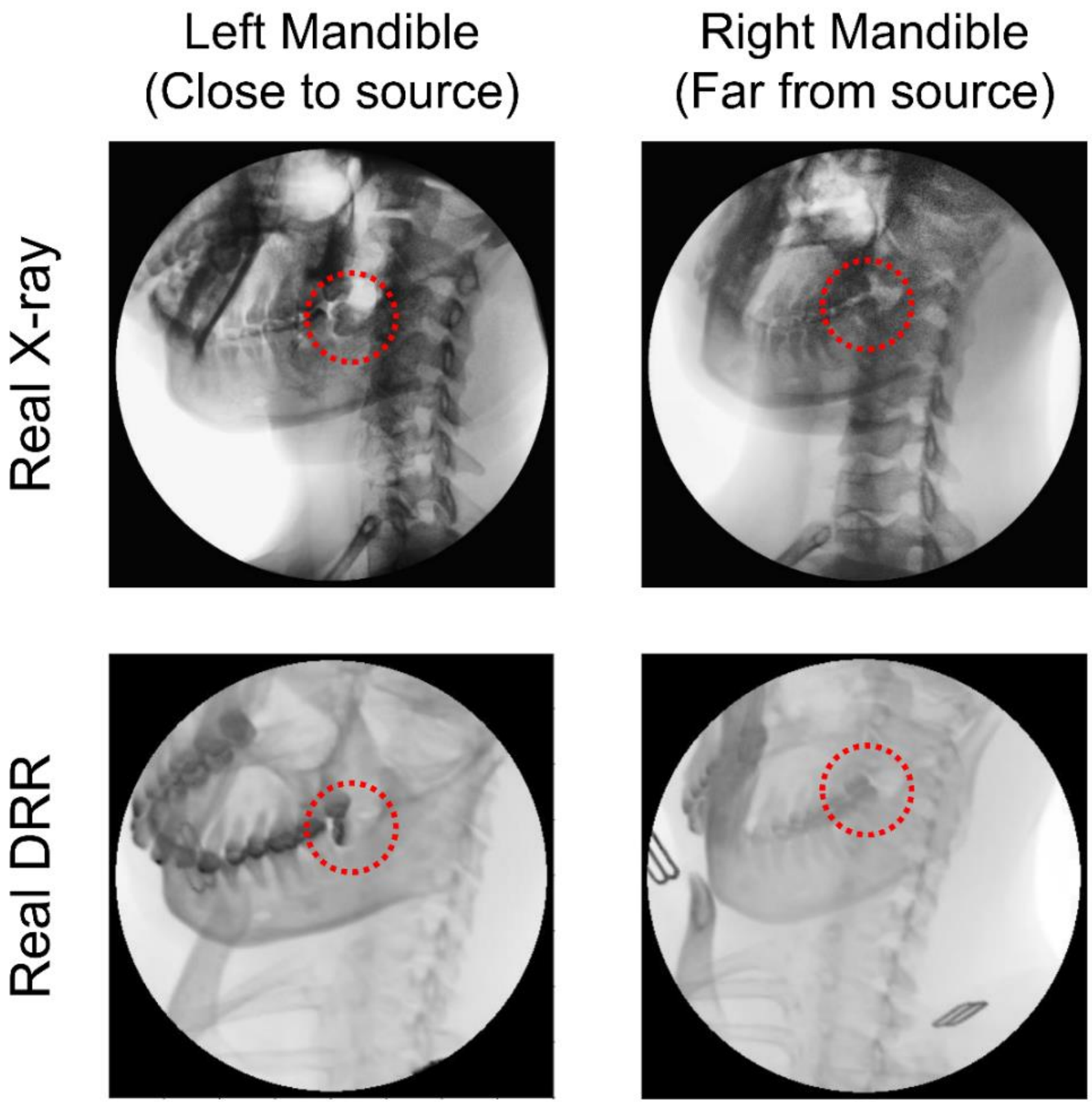

**Fig. 6**: A comparison of the renderings of real X-rays and real DRRs. The right and left wisdom teeth on the lower jaw imaged in both X-rays and DRRs were marked using *red dotted circles*.

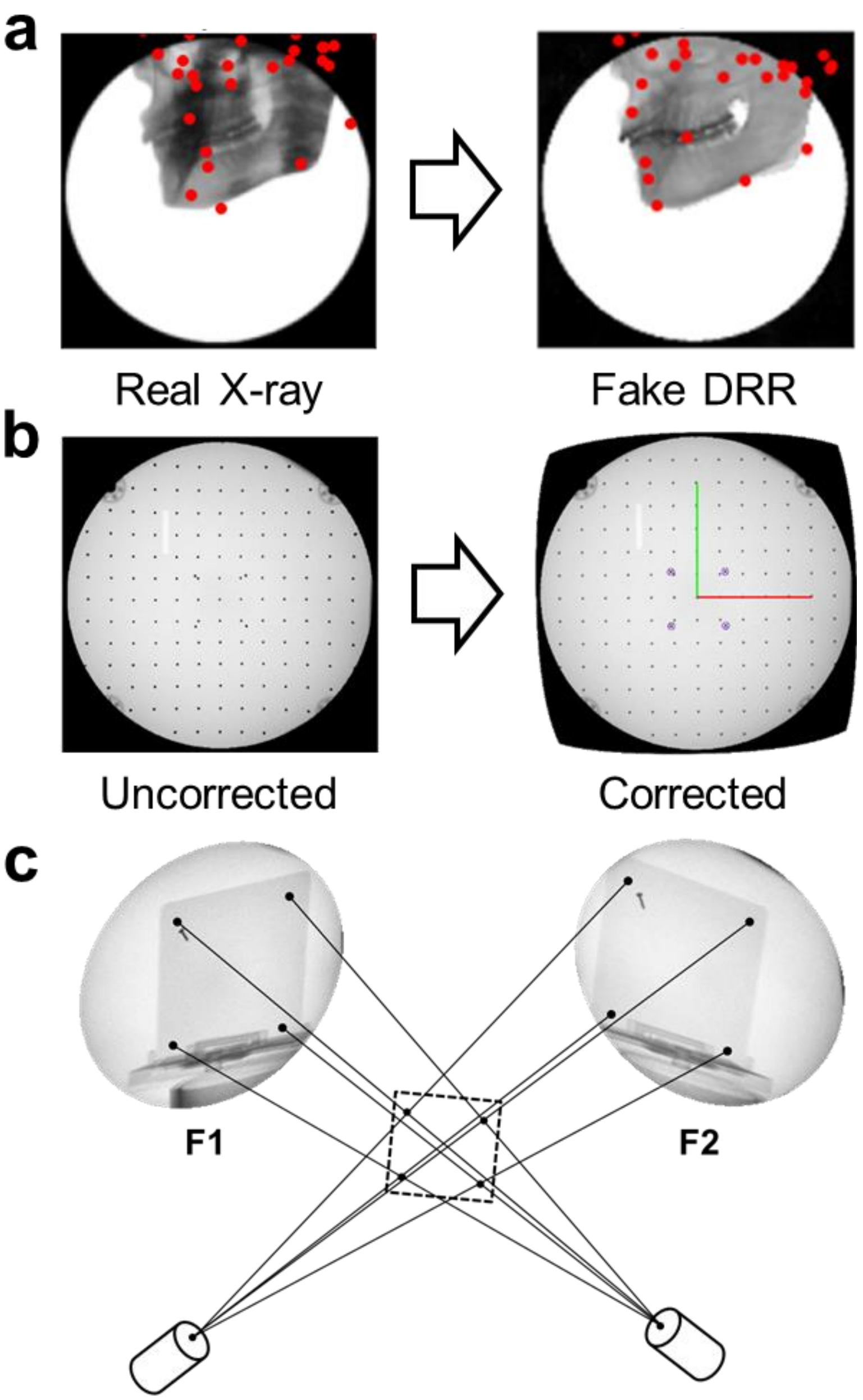

**Fig. 7**: Preprocessing of fluoroscopic X-ray images for 3D-to-2D registration. (**a**) Manual segmentation of the skull region in a real X-ray image, which were used as a mask to segment the DRR after style translation. The *red* points represent the landmarks detected using the ResNet on both the real X-ray and the corresponding DRR. (**b**) Distortion correction using an acrylic calibration plate consisting of stainless steel bead arrays. (**c**) Illustration of calibrating the relative position of dual fluoroscopic X-ray sources.

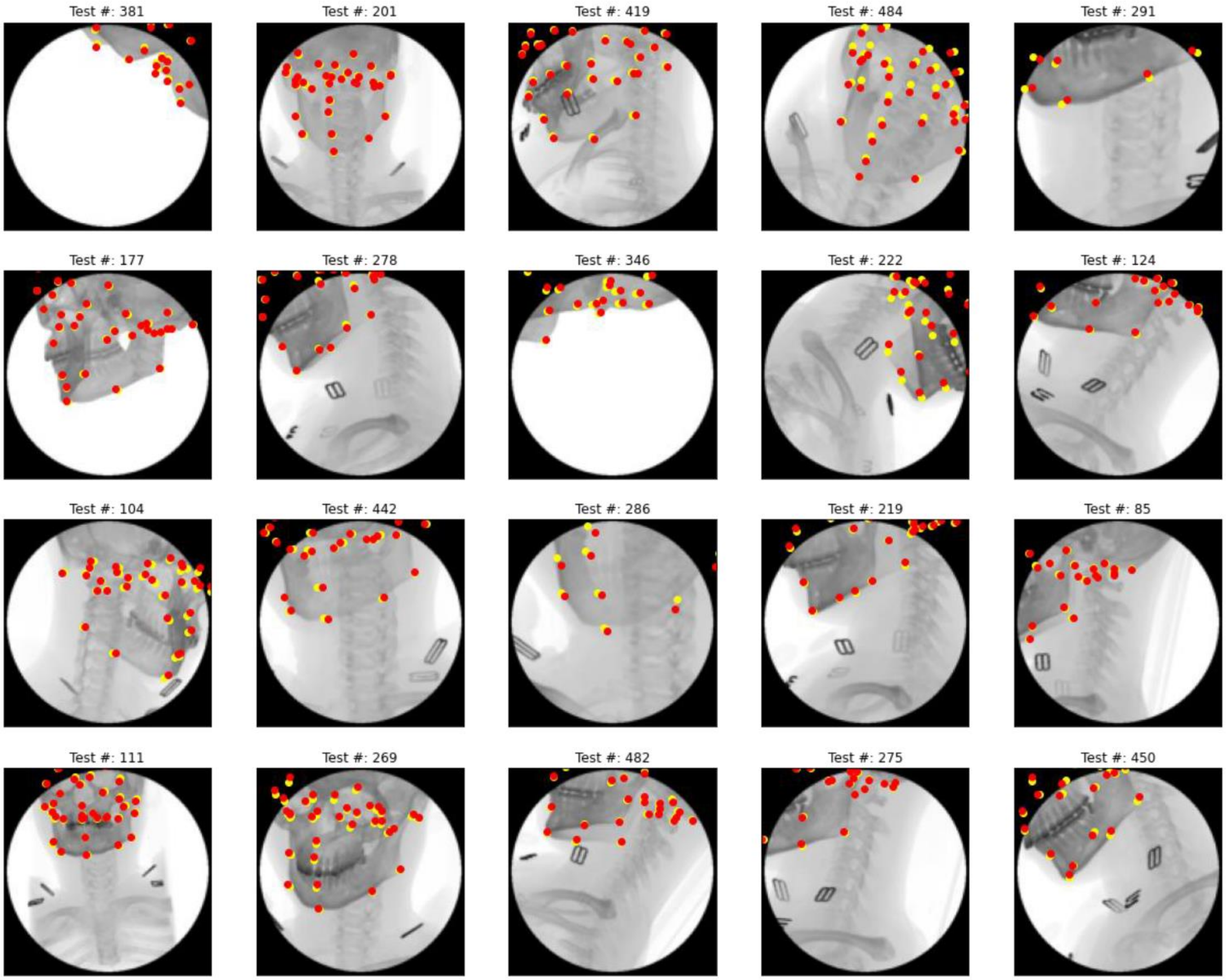

**Fig. 8**: The predictions (*red*) and labels (*yellow*) of the skull landmarks on the DRRs randomly chosen from the testing set. Note that skull-segmented DRRs were also tested (*e.g.*, images at [row, column] of [1, 1], [2, 1], and [2, 3])

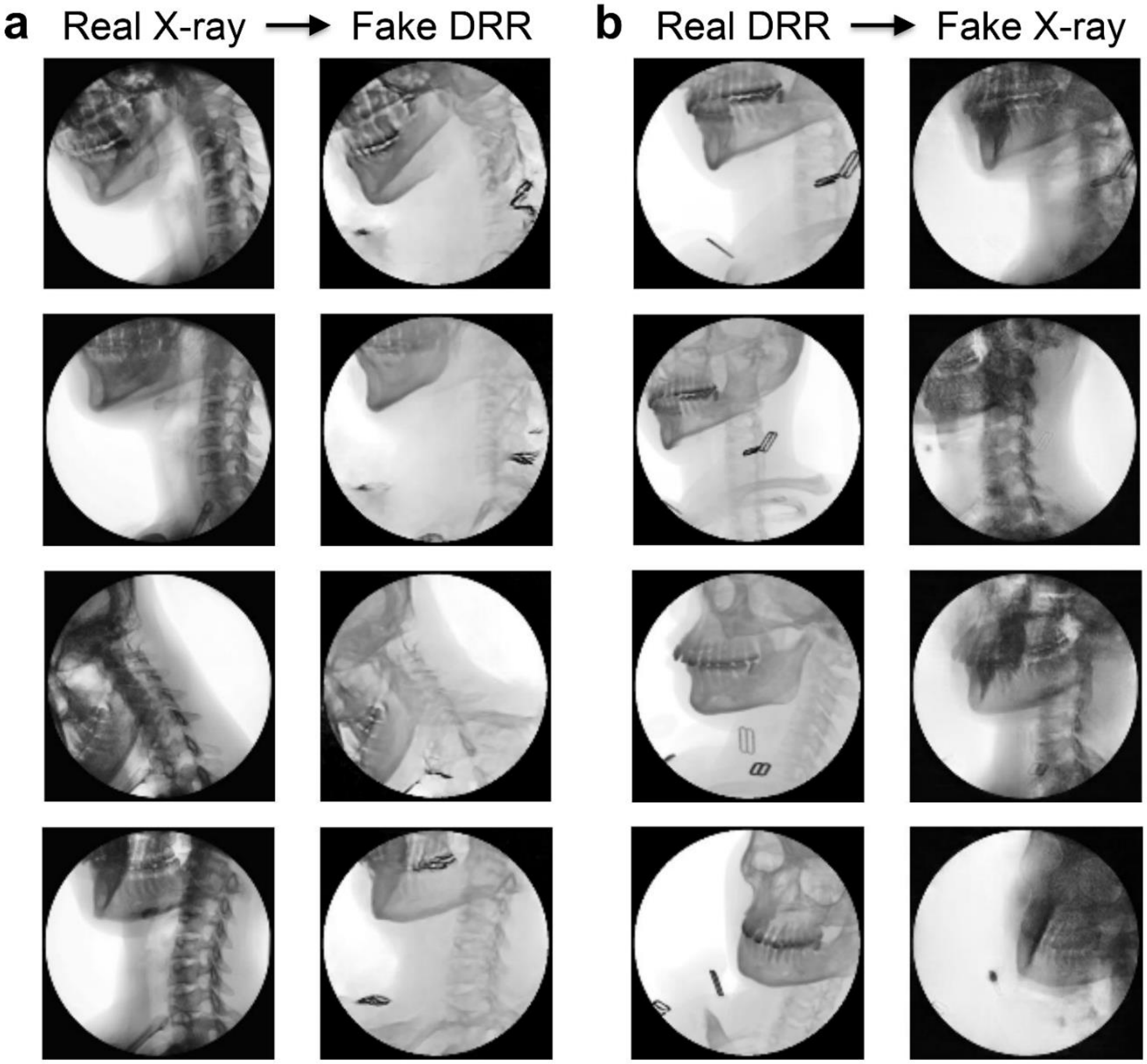

**Fig. 9**: Examples of style translations from X-rays to DRRs (**a**) and from DRRs to X-rays (**b**) using the trained cycle GAN.

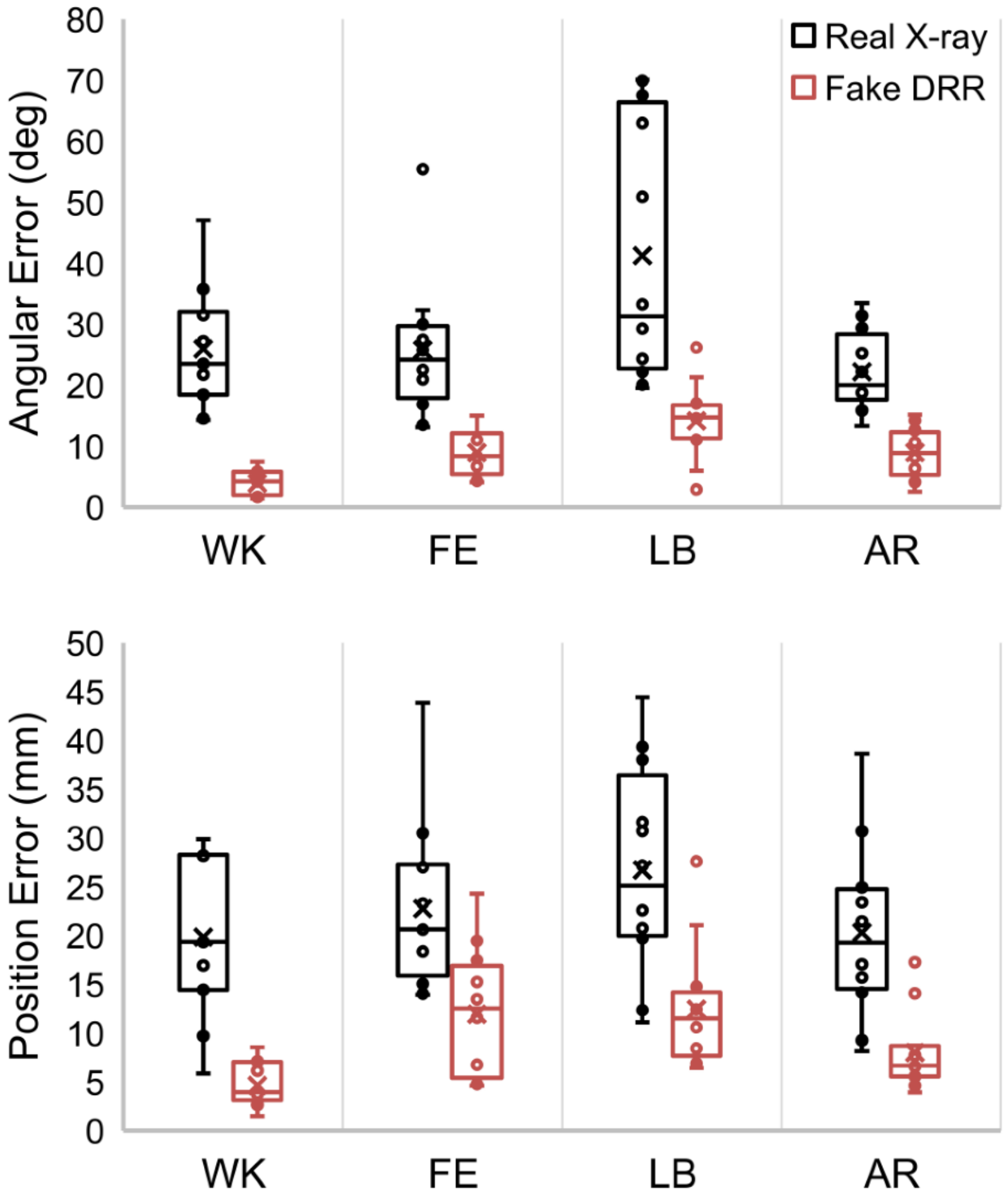

**Fig. 10**: Comparison of the 3D angular and position errors using predicted image landmarks with and without image style transfer in the registration of 3D skull modes to dual fluoroscopic images. (*WK* = walking; *FE* = flexion-extension; *LB* = lateral bending; *AR* = axial rotation).

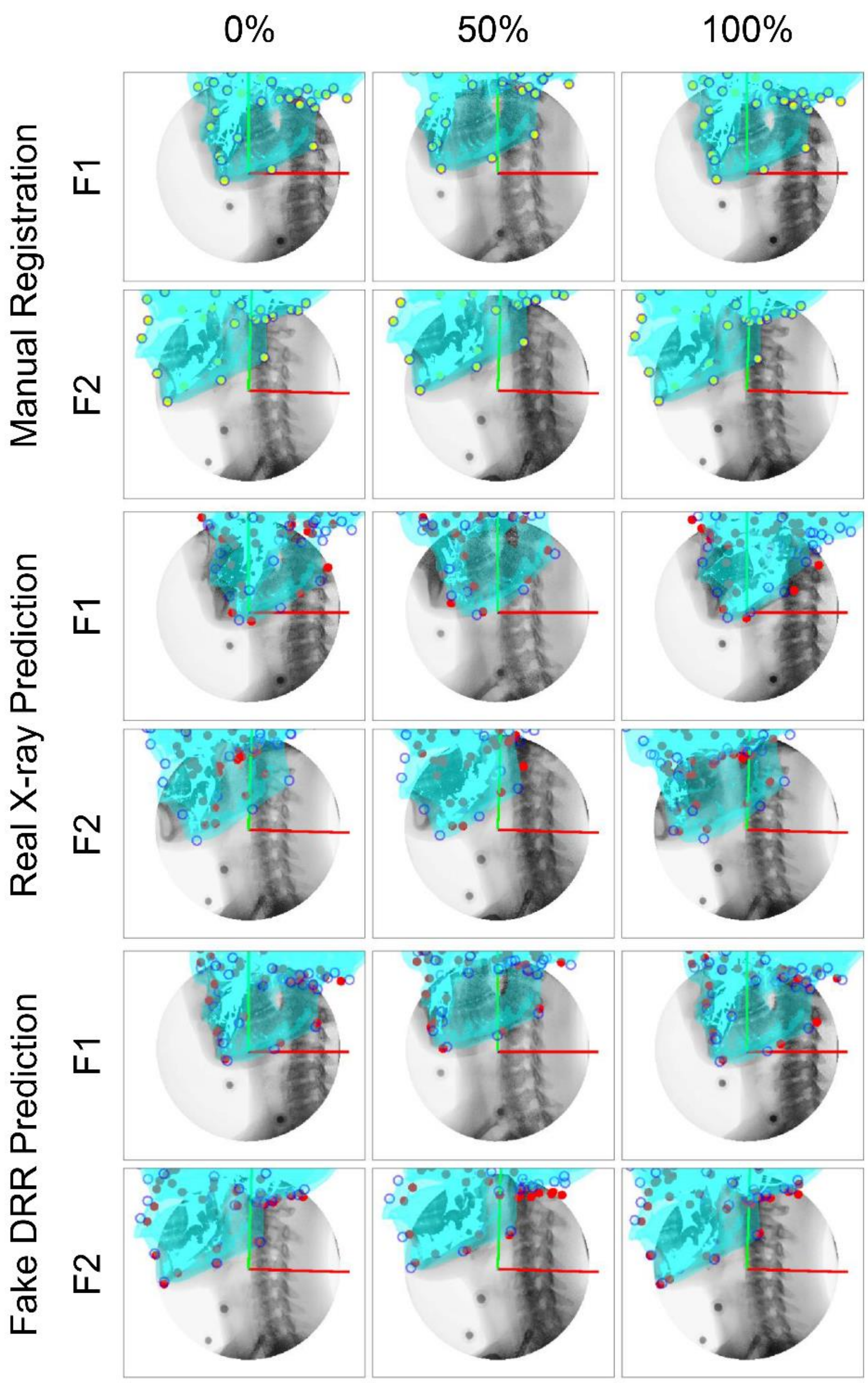

**Fig. 11**: Comparison of manual registration, real X-ray predicted registration, and fake DRR predicted registration at 0%, 50%, and 100% of a gait cycle when the subject walked on a treadmill. (*Blue cycles* = anatomic landmarks on the 3D skull model; *Yellow points* = manually registered image landmarks; *Red points* = predicted image landmarks)

## Tables

**Table 1**: The strategy to mirror predicted F1 and F2 landmarks in point-based registration.

| Optimization # | F1 Landmarks | F2 Landmarks |
|---|---|---|
| 1 | – | – |
| 2 | Mirrored | – |
| 3 | – | Mirrored |
| 4 | Mirrored | Mirrored |

**Supplementary Material**

**Appendix A: Graphic Presentation of Registration of the 3D Skull Model in Neck Functional Motions**

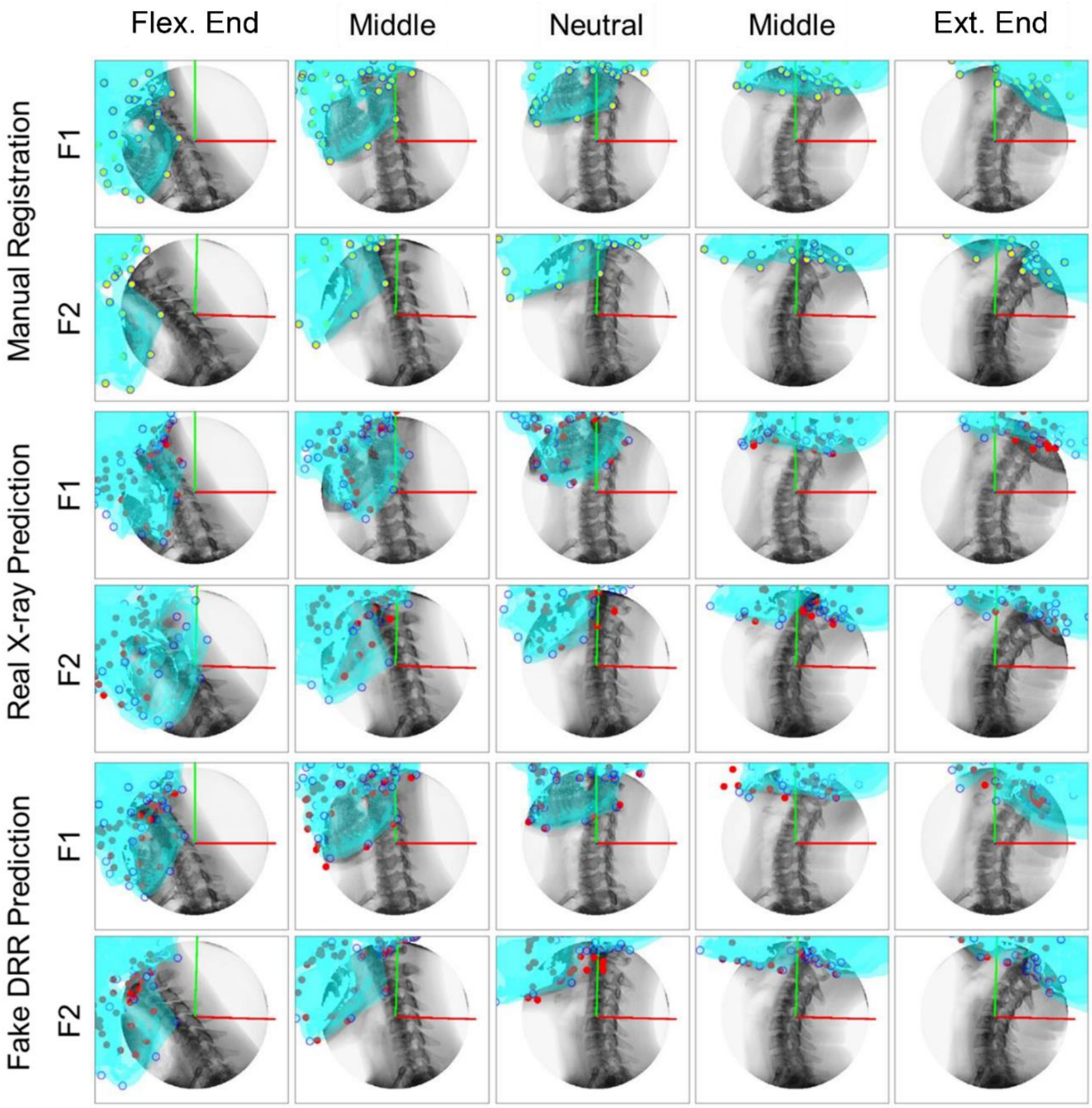

**Fig. A1**: Comparison of manual registration, real X-ray predicted registration, and fake DRR predicted registration during neck flexion and extension. (*Blue* cycles = anatomic landmarks on the 3D skull model; *Yellow* points = manually registered image landmarks; *Red* points = predicted image landmarks)

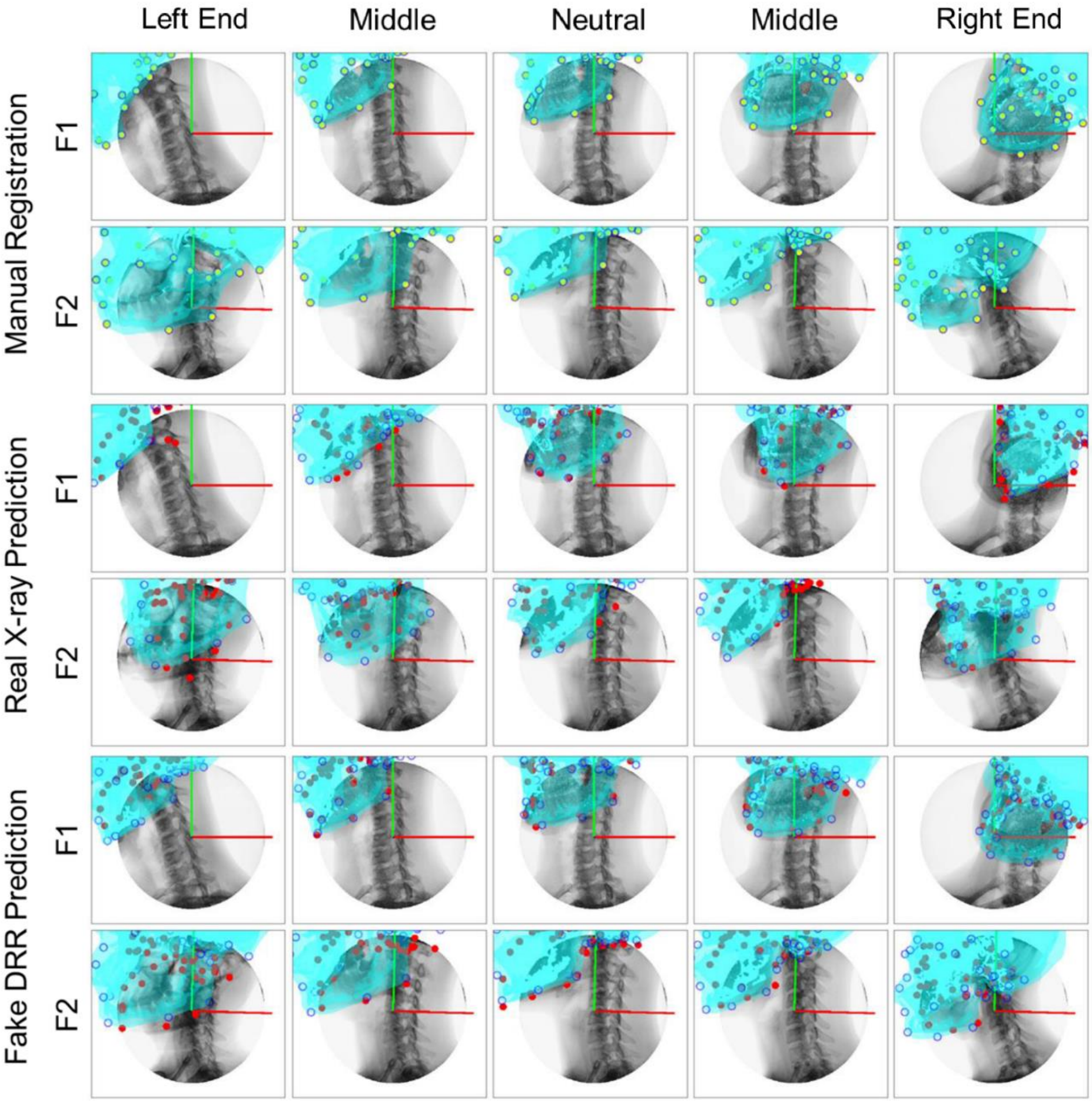

**Fig. A2**: Comparison of manual registration, real X-ray predicted registration, and fake DRR predicted registration during neck left and right lateral bending. (*Blue* cycles = anatomic landmarks on the 3D skull model; *Yellow* points = manually registered image landmarks; *Red* points = predicted image landmarks)

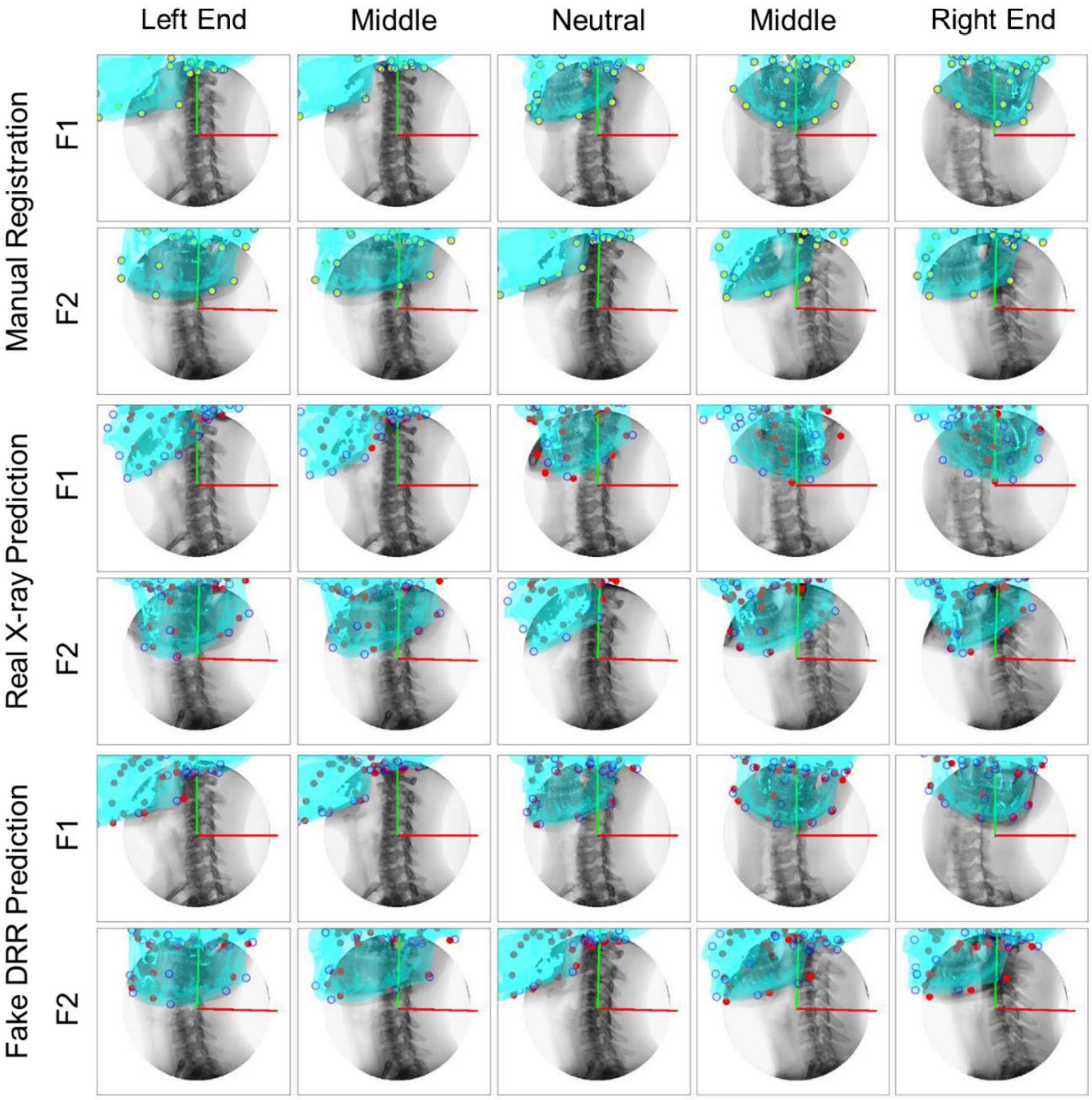

**Fig. A3**: Comparison of manual registration, real X-ray predicted registration, and fake DRR predicted registration during <u>neck left and right axial rotation</u>. (*Blue* cycles = anatomic landmarks on the 3D skull model; *Yellow* points = manually registered image landmarks; *Red* points = predicted image landmarks)